\documentclass[11pt,twoside]{book}
\usepackage{konkolyproc2}
\usepackage{longtable}
\usepackage{amsmath,amssymb}
\usepackage{graphicx}
\usepackage{lscape}
\usepackage{index}
\usepackage{natbib}
\usepackage{bigdelim}
\usepackage{multirow}
\makeindex

\begin{document}

\pagestyle{myheadings}
\setcounter{equation}{0}\setcounter{figure}{0}\setcounter{footnote}{0}\setcounter{section}{0}\setcounter{table}{0}\setcounter{page}{1}
\markboth{Szabados, Szab\'o \& Kinemuchi}{RRL2015 Conf. Papers}
\title{New systemic radial velocities of suspected RR Lyrae binary stars} 
\author{Elisabeth Guggenberger$^{1,2}$, Thomas G. Barnes$^3$ \& Katrien Kolenberg$^{4,5,6}$ }
\affil{$^1$Max Planck Institute for Solar System Research, G\"ottingen, Germany\\
$^2$Stellar Astrophysics Centre, Aarhus University, Denmark\\
$^3$McDonald Observatory, University of Texas at Austin, Texas, USA\\
$^4$University of Antwerp, Belgium\\
$^5$KU Leuven, Belgium\\
$^6$Harvard-Smithsonian Center for Astrophysics, Cambridge MA, USA}

\begin{abstract}
Among the tens of thousands of known RR Lyrae stars there are only a handful that show indications of possible binarity. The question why this is the case is still unsolved, and has recently sparked several studies dedicated to the search for additional RR Lyraes in binary systems. Such systems are particularly valuable because they might allow to constrain the stellar mass. 

Most of the recent studies, however, are based on photometry by finding a light time effect in the timings of maximum light. This approach is a very promising and successful one, but it has a major drawback: by itself, it cannot serve as a definite proof of binarity, because other phenomena such as the Blazhko effect or intrinsic period changes could lead to similar results. Spectroscopic radial velocity measurements, on the other hand, can serve as definite proof of binarity. 

We have therefore started a project to study spectroscopically RR Lyrae stars that are suspected to be binaries. We have obtained radial velocity (RV) curves with the 2.1m telescope at McDonald observatory. From these we derive systemic RVs which we will compare to previous measurements in order to find changes induced by orbital motions. We also construct templates of the RV curves that can facilitate future studies. 

We also observed the most promising RR Lyrae binary candidate, TU UMa, as no recent spectroscopic measurements were available. We present a densely covered pulsational RV curve, 
which will be used to test the predictions of the orbit models that are based on the O-C variations.
\end{abstract}

\section{Introduction}
So far our knowledge of the masses of RR Lyrae stars mainly relies on models of stellar evolution and pulsation. Having an independent measure of the mass such as the one derived from a binary orbit would be a valuable test case for our understanding of stellar pulsation. The only RR Lyrae star for which this was achieved so far turned out to be a surprise: OGLE-BLG-RRLYR-02792 turned out to have a mass of only 0.26 M$_\odot$ \citep{pietr12} and is now known as an RR Lyrae impostor which mimics the pulsation typical of RR Lyrae stars. The only other RR Lyrae star that can be considered a confirmed candidate is TU Uma with photometric and spectroscopic measurements pointing consistently towards binary motion with a period longer than 20 years \citep{wade99, liska15}. 

Concerning the large numbers of known and of well-studied RR Lyrae stars this number is surprisingly low as usually about half of all stars are considered to be part of binary or multiple systems. This lack of binaries might partly be due to the fact that not many studies have yet been dedicated to searching these objects. Several stars  have been mentioned in the literature as possible binary candidates but have not been followed up in dedicated campaigns. There are, for example, stars  for which discrepant systemic radial velocities have been measured, which was pointed out by \citet{solano97} and \citet{fern97}. These are the stars that we primarily follow up in this project. Sometimes older measurements of the center-of-mass velocity are based only on a few spectra with phases calculated from ephemeris that have later been improved. These measurements might turn out to be erroneous when revisited. With our new data we aim to verify or refute the binary nature of the objects in question.  The new accurate center-of-mass velocities can then also be used for other studies for example on galactic dynamics.

Another indication of a possible binary orbit are periodic changes in the times of maximum light, and stars with suspicious O-C diagrams (observed minus calculated time of maximum) have been included in our target list. 

Additionally, TU UMa is one of our targets. With the latest published RV value being more than a decade old, new measurements had become necessary to check the predictions from the orbit models which are based on photometry.

\section{Techniques} 

Measuring the velocity changes introduced by orbital motion is  a challenging task in RR Lyrae stars. The signal that we expect to measure is about an order of magnitude smaller than that caused by the pulsation motion of the atmospheric layers which usually amounts to tens of km/s. To derive a reliable center-of-mass velocity, one needs to understand the pulsation signal first, in order to remove it from the data. Hence several measurements, spread  over different phases of the pulsation cycle, are necessary. With periods of about half a day, this means that it is usually not possible to cover a full pulsation cycle of a target in one night. Continued observations in the subsequent nights should then not duplicate the phases, which requires careful scheduling. 

Often templates, for example that of \citet{Liu91} are used to model the RV curve of RR Lyrae stars and hence to find the center-of-mass velocity. However, when systemic RVs are derived from only a few measurements of the velocity, phases of the observations need to be known accurately (either directly from the RV measurements of from photometry), as well as the amplitude of the signal. In our project we aim at a number of 10 measurements per star in one orbit epoch, which allow for a check of the phases computed from published ephemeris.

Another challenge is the faintness of the RR Lyrae stars. This cannot be overcome by simply increasing the integration time, as the rapid atmospheric motion would then lead to smeared spectral lines. We therefore kept the integration times below 3 \% of the pulsation cycle.

\section{Observations and Reductions}
Observations were carried out with the Sandiford Echelle Spectrometer \citep{mccarthy93} in the f/13.5 Cassegrain focus of the 2.1 Otto Struve telescope at McDonald Observatory, Texas. We have continuous spectral coverage over the observed range of 4250-4750\AA~with a resolving power of R$\approx$55000. The signal to noise ratios are typically between 20 and 60, depending on the brightness of the star and the limitations on the integrations time ($<$3 \% of the pulsation period). Radial velocity standard stars were observed several times during the course of each night, and Th-Ar spectra were taken for wavelength calibration after each stellar spectrum. We used standard iraf routines for data reduction, and we derived the velocities by cross-correlating with the standard star spectra using the iraf task fxcor. Only metal lines were used in the cross-correlation, as the hydrogen lines are known to have a phase offset and different velocity amplitudes compared to iron lines.

\section{Preliminary Results and Outlook}

\begin{figure}[!ht]
\includegraphics[width=1.0\textwidth]{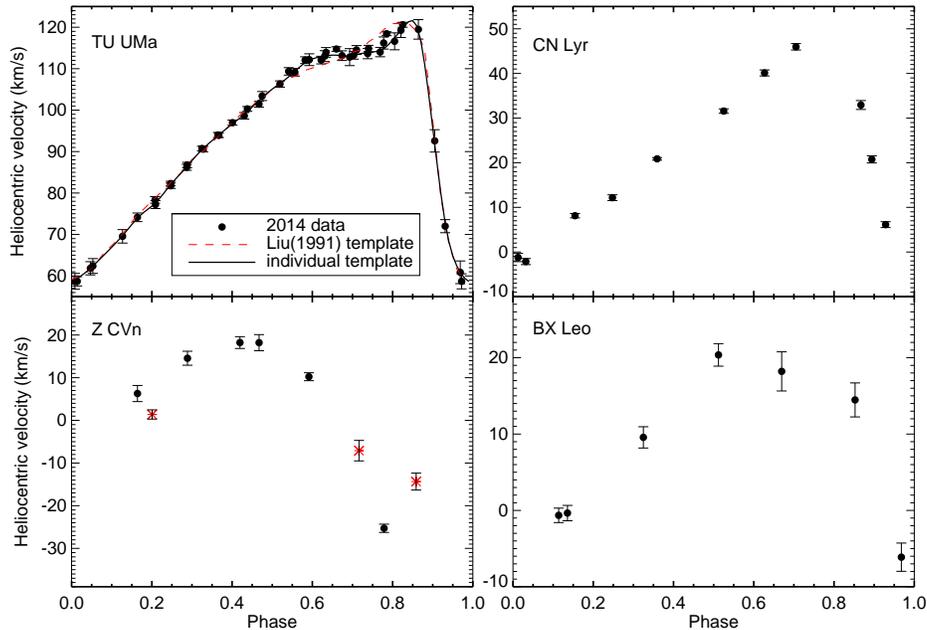}
\caption{Example RV curves obtained in the McDonald RR Lyrae binary project. TU UMa data of 2014 are shown in the upper panel, together with the scaled standard template by \citet{Liu91} (dashed red line) and the individual template created based on our data (solid black line). Other panels show the RRab star CN Lyr (upper right), the RRc star BX Leo (lower right), and the modulated star Z CVn which we observed in two different Blazhko phases which are shown as filled circles and red crosses, respectively (lower left).} 
\label{RVcurves} 
\end{figure}

During four runs between March 2014 and June 2015, 18 science objects were observed (see Table~\ref{obs}). For some of them the desired number of about 10 observations has been reached which allows the construction of an individual template. Example RV curves are shown in Fig~\ref{RVcurves}. For several others we have already obtained the minimum number of 3 observations, which allows an estimate of the systemic radial velocity based on a general template. 

The project is ongoing and we have applied for more observing time to increase the number of spectra of the stars which are not well-covered yet, as well as to observe more targets for which indications of binarity have been reported.

\begin{table}[!hb] 
\caption{Stars observed so far in the project and number of observations.}
\label{obs}
\smallskip
\begin{center}
\begin{tabular}{lclc}
\tableline
\noalign {\smallskip} 
Target & Number of spectra & Target & Number of spectra\\ 
\noalign{\smallskip}
\tableline
\noalign{\smallskip}
TU Uma & 47+8 & ST Leo & 7 \\
CN Lyr & 11 & BK And & 6 \\
DM Cyg &11 & XX Hya & 5 \\
Z CVn & 9 & RR Gem & 4 \\ 
BK Dra & 9 & RV UMa & 4 \\
AO Peg & 8 & CI And & 3 \\
AV Vir & 8 & U Tri & 2 \\
BX Leo & 7 & TT Lyn & 1 \\
SZ Leo & 7 & SS Leo & 1 \\
\noalign{\smallskip}
\tableline
\noalign{\smallskip}
\end{tabular}  \end{center}
\end{table}

The derived center-of-mass velocities will soon be published in a forthcoming paper together with the individual velocity templates to facilitate future measurements of the center-of-mass velocity. Long orbits like that of TU UMa might be common, especially as close binaries will most likely have interacted with each other or even have been destroyed during the red giant phase of the RR Lyrae star. Hence further campaigns will be necessary in the future to characterize the orbits in detail. In our study we aim to lay a solid foundation for future studies of the confirmed candidates. We therefore will also provide individual velocities of each spectrum in order to make direct comparisons possible.

\acknowledgements
The research leading to the presented results has received funding from the European Research Council under the European Community's Seventh Framework Programme (FP7/2007-2013) / ERC grant agreement no 338251 (StellarAges), from FWF grant P19962-N16 (Modulated RR Lyrae Stars) from the Austrian Wissenschaftsfonds (FWF), and from a Marie Curie International Outgoing Fellowship within FP7 (PIOF-255267, SAS-RRL).

\end{document}